\documentclass[letterpaper]{ptephy_v1}

\usepackage{subfig}
\usepackage{amsmath}
\usepackage{graphics}
\usepackage{url}
\usepackage{ulem}  

\usepackage{color}

\usepackage{enumerate}

\newcommand{\simgt}{\lower.5ex\hbox{$\; \buildrel > \over \sim \;$}}
\newcommand{\simlt}{\lower.5ex\hbox{$\; \buildrel < \over \sim \;$}}
\newcommand{\solM}{M_{\odot}}

\begin{document}

\title{The potential of advanced ground-based gravitational wave detectors to detect generic deviations from general relativity}

\author{\name{\fname{Tatsuya} \surname{Narikawa}}{\ast} and \name{\fname{Hideyuki}
\surname{Tagoshi}}{\dag}
}
\address{Graduate School of Science, Osaka City University, Sumiyoshi-ku, Osaka 558-8585, Japan
\email{narikawat@gwv.hep.osaka-cu.ac.jp}
\affil{\dag}\email{tagoshi@sci.osaka-cu.ac.jp}}

\begin{abstract}
We discuss the potential of advanced ground-based gravitational wave detectors
such as LIGO, Virgo, and KAGRA
to detect generic deviations of gravitational waveforms 
from the predictions of general relativity. 
We use the parameterized post-Einsteinian formalism to characterize the deviations, 
and assess what magnitude of deviations are detectable 
by using an approximate decision scheme based on Bayesian statistics. 
We find that there exist detectable regions of the parameterized post-Einsteinian parameters 
for different binary masses
from the observation of a single gravitational wave event. 
The regions are not excluded by currently existing binary pulsar observations
for the parameterized post-Einsteinian parameters at higher post-Newtonian order.
We also find that neglect of orbital eccentricity or tidal deformation effects do not cause a significant bias on the detectable region of generic deviations from general relativity.
\end{abstract}

\subjectindex{E02, E03}

\maketitle

\section{Introduction}
To date, general relativity (GR) has passed all experimental and observational tests 
in the weak-field and non-relativistic regimes.
Nonetheless, the following issues motivate us to study modified gravity (MG) theories:
The first one is the fundamental theoretical one that quantizes gravity and unifies it with other interactions; 
see \cite{Alexander:2007kv} for related work in the context of gravitational waves (GW).
The second one is the black hole singularity, which appears unavoidably in GR. Thus, GR is expected to be broken down at the singularity
(see, e.g.,~\cite{Nakano:2015uja} for related work in the context of GW).
The third one is the current cosmic acceleration discovered with the observation of type Ia supernovae, which may suggest the existence of dark energy~\cite{Riess:1998cb,Perlmutter:1998np}.
This is supported by the cosmic microwave background temperature anisotropies~\cite{Komatsu:2010fb}
and the large-scale structure of galaxies~\cite{Tegmark:2006az}.
However, the nature of dark energy is still unknown, and
it may also suggest a sign of breakdown of GR on a cosmologically large scale 
(see the reviews in~\cite{Clifton:2011jh,Jain:2010,Joyce:2014kja} for details).
It is thus important to experimentally test GR more precisely.
Advanced LIGO has detected two gravitational wave events from binary black hole mergers during its first observation run~\cite{Abbott:2016blz,Abbott:2016nmj,TheLIGOScientific:2016pea}. This observation suggests that advanced LIGO~\cite{Harry:2010zz} and the coming advanced Virgo~\cite{Virgo} and KAGRA~\cite{Somiya:2011np,Aso:2013eba} will observe a lot of similar GW events when these detectors are in full operation.  These observations will give us a great opportunity to test GR through comparison of observed gravitational waveforms with the predictions~\cite{TheLIGOScientific:2016src}.
We might encounter breakdown of GR in strong-field and dynamical regimes in the coming GW observations.
They would provide us with lots of important clues to construct a correct theory of gravity 
that overcomes the above problems.
It is thus important to quantify what deviations from GR can be detected by the coming GW observations.

The second-generation ground-based laser interferometer is sensitive to gravitational waves 
in the frequency band between $10$ and a few $1000~{\rm Hz}$.
The GWs emitted during the inspiral phase of coalescing compact binary (CCB) systems
are one of the most promising sources for these detectors,
which will be able to detect the GWs from CCB systems, composed of neutron stars (NSs) 
and/or black holes (BHs), within a few $100~{\rm Mpc}$ or a few ${\rm Gpc}$.

The inspiral phase of CCB systems is well understood thanks to 
the post-Newtonian (PN) formalism~\cite{Blanchet:2006zz,Blanchet:2013haa}.
Therefore, GW observations of the inspiral signals emitted from CCB systems can be a powerful probe of 
the strong-field, dynamical aspect of gravity theories~\cite{Yunes:2013dva}.

It is extremely instructive to contrast predictions of parameterized deviations from GR. 
For example, the parameterized post-Newtonian (ppN) formalism has been used for weak-field tests of gravity~\cite{Will:2014kxa}.
Many previous works take a model-independent approach to the strong-field test of gravity for the coming GW observations
(see the reviews in~\cite{Yunes:2013dva,Gair:2012nm,Berti:2015itd,Broeck:2013kx} of GW tests of gravity).
In the PN formalism, the amplitude and the orbital phase of the inspiral signal are expanded 
in powers of $v/c$, with $v$ a characteristic velocity~\cite{Blanchet:2006zz,Blanchet:2013haa}.
A test of GR could take the form of modifying the PN phasing coefficients, 
called parameterized tests of post-Newtonian theory (PTPN)~\cite{Arun:2006yw,Arun:2006hn,Mishra:2010tp}.
A general algorithm for testing GR, called TIGER (Test Infrastructure for GEneral Relativity), was developed, based on Bayesian model selection~\cite{Li:2011cg,Li:2011vx,Broeck:2013kx,Agathos:2013upa}.
Given the data containing the inspiral signal, one compares the Bayesian evidence
for the GR hypothesis against the MG hypothesis in which one or more phasing coefficients in the PN formula are modified.

Yunes and Pretorius have developed the parameterized post-Einsteinian (ppE) formalism, 
which is a similar but more general approach than the PTPN, covering a large class of 
parameterized deviations~\cite{Yunes:2009ke}.
In~\cite{Yunes:2010qb}, the authors have shown that lower-PN corrections in the phase have been 
strongly constrained with the observation of the orbital period decay of the binary pulsar ${\rm PSR~J0737-3039}$. 
In~\cite{Cornish:2011ys}, the authors have evaluated the parameter estimation accuracy of the ppE phasing  parameters 
for observations with Advanced LIGO/Advanced Virgo and LISA detections
by using the Markov chain Monte Carlo method. 
Their results have shown that GW constraints are stronger than those of binary pulsar observations at higher PN order.
There have been many other works which used the ppE formalism~\cite{Yunes:2010qb,Chatziioannou:2012rf,Vallisneri:2013rc,Sampson:2013lpa,Sampson:2013wia,Sampson:2013jpa,Loutrel:2014vja,Tso:2015vua}.

To evaluate the parameter estimation accuracy, Fisher information matrix analysis has often been used~(see, e.g.,~\cite{Cutler:1994ys,Poisson:1995ef}).
Bayesian hypothesis testing is also useful for model selection in the GW data analysis~(e.g.,~\cite{DelPozzo:2011pg}).
Recently, Vallisneri has proposed a simple approximate decision scheme 
for testing MG within the framework of Bayesian hypothesis testing~\cite{Vallisneri:2012qq}. 
In this approach, one can easily evaluate the signal-to-noise ratio (SNR) 
required for detection of a given deviation from GR waveforms.
In this method, one only needs to compute the SNR and a fitting factor (FF), which are computationally cheaper than computing the full evidence\footnote{Note that this approximate scheme is valid only in the limit of large SNR 
and small deviations from GR waveforms. See Sect.~\ref{sec:statistics} for the detail of this scheme.}.
Vallisneri's method is indeed useful to assess the detectability of MG signals. 
We have already demonstrated by using this method that there exist detectable regions of the bigravity parameters (the graviton mass and 
the ratio between gravitational constants for two kinds of graviton), 
which characterize graviton oscillations, 
for a single observation of GW from CCB with an advanced ground-based GW detector~\cite{Narikawa:2014fua}. 

For detection purposes, one usually uses a GR waveform as a template for the matched filtering analysis. 
However, if GR happened to be broken in the strong-field regime, 
GR templates could still extract signals, but with the wrong values of the parameters. 
This is called fundamental bias~\cite{Yunes:2009ke}\footnote{Fundamental bias is systematic error due to the use of waveforms derived in the incorrect theory.}.
If the deviations from GR are not large enough to be detectable and 
yet this systematic error is larger than the statistical uncertainties in parameter estimation, 
fundamental bias can be hidden, and becomes stealth bias.
In~\cite{Vallisneri:2013rc}, Vallisneri and Yunes have investigated the stealth bias by using 
the Vallisneri's method and
have indicated that stealth bias could be a generic feature of GW observation, 
if GR is not correct in the strong-field regime.

In this paper, looking toward the full operation of advanced ground-based GW detectors,
we explore the detectability of generic deviations of gravitational waveforms from CCBs
by using the ppE formalism based on Vallisneri's method.
We consider non-spinning binary systems consisting of binary neutron stars (BNS) , neutron star$-$black hole (NSBH) binaries, and binary black holes (BBHs).
We explore what deviations from GR waveforms are detectable.
We also focus on what the effect of systematic bias is on the detectable region.
We investigate the systematic bias~\cite{Favata:2013rwa} due to the neglect of spin~\cite{Arun:2008kb}, orbital eccentricity~\cite{Favata:2013rwa,Krolak:1995md}, and tidal deformation~\cite{Flanagan:2007ix,Vines:2011ud,Damour:2012yf,Agathos:2015uaa} on constraining generic deviations from GR waveforms. 
Bayesian analysis has been performed~\cite{Agathos:2013upa,Sampson:2013lpa} on testing GR with GWs including spins and the NS finite‐size effect.
The authors of \cite{Agathos:2013upa} have studied the robustness of TIGER against unknown fundamental, astrophysical, and instrumental effects.
They concluded that the $400~{\rm Hz}$ cutoff renders tidal effects up to 6PN order in phase invisible without affecting TIGER's ability to look for GR violations. 
The authors of \cite{Sampson:2013lpa} have studied the effect of aligned spins in the templates on 
the Bayes factors between a 1PN ppE model and GR 
in the case when the signal is a ppE model without spin.
They have shown that the degeneracy between the 1.5PN spin term and the 1PN ppE parameter significantly weakens the bounds.
In this paper, we include higher PN order terms of spins, tidal deformations, and orbital eccentricity, and study 
the case when both signals and templates include these additional parameters.

The remainder of this paper is organized as follows.
In Sect. \ref{sec:ppE}, we briefly review the ppE formalism
and PN phase including corrections by spin, orbital eccentricity, and tidal deformation.
In Sect. \ref{sec:statistics}, we briefly review GW data analysis for CCBs and Vallisneri's method.
In Sect. \ref{sec:result}, we show the detectable region of the ppE parameters.
Section~\ref{sec:summary} is devoted to summary and discussion.
%

\section{The ppE formalism as generic deviations from general relativity}
\label{sec:ppE}
In this section, we briefly review the parameterized post-Einsteinian (ppE) waveform, 
which has been proposed by Arun {\it et al.}~\cite{Arun:2006yw} and Yunes and Pretorius~\cite{Yunes:2009ke}.

We consider the quasi-circular inspiral phase of GWs from CCB systems. 
In this paper, we consider non-spinning binaries for simplicity.  
The gravitational waveforms received 
by the laser interferometer are described by using 
two polarizations of GWs \footnote{For simplicity, 
we do not consider additional polarizations, which are discussed in the context of the extended ppE formalism 
in~\cite{Chatziioannou:2012rf}.}
and the detector’s antenna pattern function, which depends on the location of the source on the sky
and the detector's location. 
Since we consider detection with a single detector, 
we only need simplified forms of the waveform in which 
the detector's antenna pattern function and the inclination angle of the orbital plane are
contained in the constant amplitude and phase. 
The inspiral waveform in GR is described in the frequency domain as 
\begin{eqnarray}
 \tilde{h}_{\rm GR}(f) = A_{\rm GR}(f) e^{i \Psi_{\rm GR}(f)}~,
\end{eqnarray}
where the amplitude $A_{\rm GR}(f)$ (up to Newtonian order) and the phase function 
$\Psi_{\rm GR}(f)$ are given as a series in the inspiral reduced frequency $u \equiv (\pi {\cal M} f)^{1/3}$ as,
\begin{eqnarray}
 A_{\rm GR}(f) = {\cal A}\frac{{\cal M}^2}{D_{\rm L}} u^{-7/2},
\end{eqnarray}
and
\begin{eqnarray}
 \Psi_{\rm GR}(f) = 2 \pi f t_{\rm c} + \Phi_{0} -\frac{\pi}{4} + \frac{3}{128 u^{5}} \sum_{k=0}^7 
 \left( c_k^{\rm PN} + l_k^{\rm PN} \ln u \right) u^{k},
\end{eqnarray}
where ${\cal A}$ is a constant determined by the sky location and the inclination angle,
$D_{\rm L}$ is the luminosity distance to the source,
$t_{\rm c}$ is the coalescence time, and $\Phi_{0}$ is a constant phase term. 
$\cal M$ is the chirp mass defined by the component masses, $m_{\rm 1}$ and $m_{\rm 2}$,
as $(m_1 m_2)^{3/5}/(m_1+m_2)^{1/5}$.
The coefficients $c_k^{\rm PN}$ and $l_k^{\rm PN}$ depend on 
the component masses $m_{\rm 1}$, $m_{\rm 2}$.

The ppE formalism has been formed by adding amplitude and phase corrections to 
the GR response function.
The ppE inspiral waveform has the form~\cite{Yunes:2009ke}
\begin{eqnarray}
 \tilde{h}_{\rm MG}(f) = A_{\rm MG}(f) e^{i \Psi_{\rm MG}(f)}~,
\end{eqnarray}
where
\begin{eqnarray}
A_{\rm MG}(f) & = & \left( 1+\alpha u^a \right) A_{\rm GR}(f), \label{hppEamp} \\
\Psi_{\rm MG}(f) & = & \Psi_{\rm GR}(f) + \beta u^b. \label{hppEphase}
\end{eqnarray}
The ppE formalism is corrected by ppE amplitude and phase functions 
with free magnitude parameter $\alpha$ and powers of the frequency $a$ in the amplitude, 
and free magnitude parameter $\beta$ and powers of the frequency $b$ in the phase.
The $\alpha u^a$ term in the amplitude corresponds to 
relative $a/2$ PN order 
with respect to the leading GR terms,
and the $\beta u^b$ term corresponds to 
relative $(b+5)/2$ PN order, respectively. 
GR is recovered when $(\alpha,~\beta) = (0,~0)$.
We restrict $a$ and $b$ to integer values since we consider circular orbits.
\footnote{However, non-integer powers 
of the frequency in the phase of the GWs arise for the non-circular orbit~\cite{Chatziioannou:2012rf}.}
In this paper, we consider the following two cases:
\begin{enumerate}[A.]
\item ~Amplitude corrections with $a \in [ -2, ~6 ]$ 
corresponding to $-1$ to 3~PN order.
\item ~Phase corrections with $b \in [ -7, ~2 ]$
corresponding to  $-1$ to 3.5~PN order. 
\end{enumerate}
Notice that for amplitude corrections, 0~PN order corresponds to quadrupole radiation, which is the dominant mode for GR,
$-1$~PN order is a typical signature of dipole radiation, and higher PN order terms correspond to higher harmonics.
We take the following parameters as independent parameters for GR:
$\theta_{\rm GR}=\{M_t, ~q, ~t_c, ~\Phi_{0} \}$, where $M_t$ is the total mass and $q$ is the mass ratio.
\footnote{We fix the distance to the source $D_{\rm L}$ and we assume a signal from a face-on binary system at the zenith, for simplicity.}
On the other hand, we take $\theta_{\rm MG}=\{ a,~\log\alpha,~\theta_{\rm GR} \}$ and 
$\theta_{\rm MG}=\{ b,~\log\beta,~\theta_{\rm GR} \}$ as independent parameters for cases A and B, respectively.\footnote{
We are using the ppE formalism with a single non‐GR parameter in the amplitude and phase. Such a simple ppE template is sufficient in terms of detecting deviations from GR and higher order corrections are irrelevant, as was found in Ref.~\cite{Sampson:2013lpa}
}.

Systematic biases caused by approximating waveforms can affect detectable regions of the ppE deviations from GR.
To study how significant these physical effects in CCB systems are,
we use a modified gravity signal with, e.g., non‐vanishing spins and recover this with the GR non‐spinning templates.
We sum the several contributions to the phase as
\begin{eqnarray}
 \Psi_{\rm GR}(f) \rightarrow \Psi_{\rm GR}(f) 
 + \Delta\Psi^{\rm spin}_{\rm 3PN}
 + \Delta\Psi^{\rm ecc.}_{\rm 2PN}
 + \Delta\Psi^{\rm tidal}_{\rm 7.5PN}~
 \label{phase_sys_GR}
\end{eqnarray}
for GR.
On the other hand, we take
\begin{eqnarray}
 \Psi_{\rm MG}(f) \rightarrow \Psi_{\rm MG}(f) 
 + \Delta\Psi^{\rm spin}_{\rm 3PN}
 + \Delta\Psi^{\rm ecc.}_{\rm 2PN}
 + \Delta\Psi^{\rm tidal}_{\rm 7.5PN}~
 \label{phase_sys_ppE}
\end{eqnarray}
for ppE phase. 
Here, 
$\Delta\Psi^{\rm spin}_{\rm 3PN}$ is the non-precessing spin effect up to the 3PN order relative to the Newtonian 0PN term in the phase~\cite{Favata:2013rwa}.
$\Delta\Psi^{\rm ecc.}_{\rm 2PN}$ is the orbital eccentricity corrections to the 2PN order relative to the leading term~\cite{Favata:2013rwa}, with the orbital eccentricity $e_0$ at a reference frequency $f_0=10~{\rm Hz}$.
The leading order of the tidal correction is relative 5PN order to the Newtonian 0PN term in the phase~\cite{Flanagan:2007ix}, but the prefactors, the tidal deformability parameter $\lambda$ is not small;
$\hat{\lambda} \equiv \lambda/M_t^5 
\sim 5 - 44$ for BNS with $(1.4~\solM,~1.4~\solM)$~\cite{Damour:2012yf}.\footnote{$\lambda$ depends on the NS masses and equation of state.}
The effects of tidal deformation have been derived up to the relative 2.5PN order to the leading term (or 7.5PN in the phase) $\Delta\Psi^{\rm tidal}_{\rm 7.5PN}$~\cite{Agathos:2015uaa}.
(Here we assume that $\lambda$ is same value for both binary components.)
In this work we study the effect of each physical correction in phases (\ref{phase_sys_GR}) and (\ref{phase_sys_ppE}).

\section{Decision scheme to assess detectability of modified gravity}
\label{sec:statistics}
In this section, we briefly review the GW data analysis and an approximate decision scheme 
to assess the detectability of the MG effects on waveforms. 
%

First, we define the noise-weighted inner product $( \hspace{-.3em}~\cdot \hspace{-.3em} ~| \hspace{-.3em} ~\cdot \hspace{-.3em} ~)$ for signals $h_{\rm A}$ and $h_{\rm B}$ as
\begin{eqnarray}
 (h_{\rm A}|h_{\rm B}) \equiv 4{\rm Re}\int_{f_{\rm min}}^{f_{\rm max}} 
 \frac{\tilde{h}_{\rm A}(f) \tilde{h}_{\rm B}(f)}{S_n(f)}df, \label{inner_product}
\end{eqnarray}
where $S_n(f)$ is the one-sided noise power spectrum density of a detector.
In this paper, we use the noise power spectrum density of the advanced LIGO that is called 
Zero Det, High Power~\cite{AdvLIGOZDHP}.
The interval of integration $f_{\rm min}$ and $f_{\rm max}$ is taken to be 
$f_{\rm min}=f_{\rm low}$ and $f_{\rm max}=f_{\rm ISCO} \approx (6^{3/2} \pi M_t)^{-1}$,
where $f_{\rm low}$ is the lower cutoff frequency that is defined for each detector and 
we take it to be $f_{\rm low}=20~{\rm Hz}$,
while $f_{\rm ISCO}$ is the frequency at the innermost stable circular orbit of the binary.

The signal-to-noise ratio (SNR) for a given signal $h$ is its norm defined as
\begin{eqnarray}
 {\rm SNR} \equiv |h| = \sqrt{(h|h)}.
\end{eqnarray}
We also define the fitting factor (FF) \cite{Apostolatos:1995pj} 
that is used to characterize the deviations of MG waveforms 
from the GR waveforms. The FF between the GR and MG waveforms is defined as
\begin{eqnarray}
 {\rm FF}(\theta_{\rm MG}) \equiv \max_{\theta_{\rm GR}}
 \frac{(h_{\rm GR}(\theta_{\rm GR})|h_{\rm MG}(\theta_{\rm MG}))}
 {|h_{\rm GR}(\theta_{\rm GR})||h_{\rm MG}(\theta_{\rm MG})|}, \label{FF}
\end{eqnarray}
where $h_{\rm GR} (\theta_{\rm GR})$ and $h_{\rm MG}(\theta_{\rm MG})$ are the GR and MG waveforms.
By definition, the maximum of FF is 1, which is realized when the MG waveform coincides with the GR waveform. 
Thus, $1-{\rm FF}$ measures the strength of the MG corrections
that cannot be absorbed in the GR waveform even if the source parameters of the GR waveform are changed.

Next, let us briefly review Bayesian hypothesis testing.
Bayes' theorem states that the 
the posterior probability distribution of the hypothesis ${\cal H}$ given the data $s$, 
$p( \vec{\theta} | s, {\cal H} )$, is given as
\begin{eqnarray}
 p( \vec{\theta} | s, {\cal H} ) = \frac{ p( \vec{\theta} | {\cal H} ) p(s | \vec{\theta}, {\cal H} ) }{ p( s | {\cal H} ) },
 \label{Bayestheorem}
\end{eqnarray}
where $p(s | \vec{\theta}, {\cal H} )$ is the likelihood function for the observation $s$, assuming the hypothesis 
${\cal H}$ and given values of the parameters $\vec{\theta}$. 
$p( \vec{\theta} | {\cal H} )$ is the prior probability distribution of the unknown parameter vector $\vec{\theta}$ 
within the hypothesis ${\cal H}$. 
$p( s | {\cal H} )$ is the fully marginalized likelihood or evidence for ${\cal H}$ defined as
\begin{eqnarray}
 p(s | {\cal H} ) \equiv \int d\vec{\theta} p( s | \vec{\theta}, {\cal H} ) p( \vec{\theta} | {\cal H} ),
\end{eqnarray}
which is the integral of the likelihood $p(s | \vec{\theta}, {\cal H} )$ multiplied by the prior over all parameters 
within the hypothesis ${\cal H}$. This is the normalization constant in the denominator of Eq.~(\ref{Bayestheorem})
for the hypothesis ${\cal H}$.
Here the data $s$ obtained from the detector is assumed to be the sum of the detector noise $n$ and the inspiral signal $h$,
\begin{eqnarray}
 s(t) = n(t) + h( \vec{\theta}; t ).
\end{eqnarray}
We assume that the noise is stationary and Gaussian. The probability distribution of the noise 
is given as
\begin{eqnarray}
 p[n] = {\cal N} e^{-(n|n)/2},
\end{eqnarray}
where ${\cal N}$ is a normalization factor.
Then, the likelihood function is simply the Gaussian distribution
\begin{eqnarray}
 p(s | \vec{\theta}, {\cal H} ) = {\cal N} e^{-( s-h(\vec{\theta}) | s-h(\vec{\theta}) )/2}.
\end{eqnarray}

Bayesian model selection is performed by comparing the posterior probabilities for different hypotheses. 
The Bayesian odds ratio for MG over GR is defined as
\begin{eqnarray}
 {\cal O} \equiv \frac{P({\rm MG}|s)}{P({\rm GR}|s)}
 =\frac{P({\rm MG})}{P({\rm GR})} \frac{P(s|{\rm MG})}{P(s|{\rm GR})},
\end{eqnarray}
where $P({\rm MG}|s)$ and $P({\rm GR}|s)$ are the posterior probabilities of the MG and 
the GR hypotheses for given data $s$,
$P({\rm MG})$ and $P({\rm GR})$ are the prior probabilities of the MG and the GR 
hypotheses, and $P(s|{\rm MG})$ and $P(s|{\rm GR})$ are the evidence of the MG and GR hypotheses.
Here, the ratio between different evidences ${\rm BF}=P(s|{\rm MG}) / P(s|{\rm GR})$ is the Bayes factor between 
two competing hypotheses, MG and GR, which can assess how much more likely the data $s$ is 
under MG rather than under GR.

Following~\cite{Vallisneri:2012qq,Vallisneri:2013rc}, we explain Vallisneri's method based on Bayesian hypothesis testing.
In Bayesian statistics, one declares the detection of MG corrections to GR when the odds ratio  exceeds a chosen threshold ${\cal O}_{\rm thr}$. 
We define ${\cal O}_{\rm MG}$ as the odds ratio when the data contain an MG signal,
with ${\cal O}_{\rm GR}$ as the one when the data contain a GR signal.
The distribution of ${\cal O}_{\rm GR}$ determines the background of false MG detections 
for a chosen threshold ${\cal O}_{\rm thr}$.
We set the threshold ${\cal O}_{\rm thr}$ by requiring a sufficiently small false alarm probability, 
$P_F=P({\cal O}_{\rm GR}>{\cal O}_{\rm thr})$, 
which is the fraction of observations in which the odds ratio ${\cal O}_{\rm GR}$ happens to exceed 
${\cal O}_{\rm thr}$ for the GR signal.
On the other hand,
the true detection probability (also known as the efficiency of the detection) $P_E=P({\cal O}_{\rm MG}>{\cal O}_{\rm thr})$ is the fraction of observations in which the odds ratio ${\cal O}_{\rm MG}$ exceeds ${\cal O}_{\rm thr}$ for the MG signal. 

We use a simple approximation for the Bayesian odds ratio. 
The logarithm of the Bayesian odds ratio scales as 
${\rm SNR}^2(1-{\rm FF})$ in the limit of large SNR and small MG deviations from GR waveforms.
\footnote{This approximation was first pointed out by Cornish {\it et al}.~\cite{Cornish:2011ys}.
Vallisneri derived a similar approximation from the Fisher information matrix.
More recently, Del Pozzo {\it et al}.~\cite{DelPozzo:2014cla} compared the prediction 
from this approximation for the Bayesian odds ratio against numerical simulation.
They found that this approximation recovers the numerical result with good accuracy when the FF value is close to unity.}
When the data contain a GR signal, we have
\begin{eqnarray}
 {\cal O}_{\rm GR} = {\cal N} e^{{x}^2/2}, \label{oddsGR}
\end{eqnarray} 
and when the data contain an MG signal, we have
\begin{eqnarray}
 {\cal O}_{\rm MG} = {\cal N} e^{{x}^2/2 + \sqrt{2}x{\rm SNR}_{\rm res} + {\rm SNR}_{\rm res}^2}, \label{oddsMG}
\end{eqnarray} 
where the residual signal-to-noise ratio, ${\rm SNR}_{\rm res}$,
is defined as
\begin{eqnarray} 
 {\rm SNR}_{\rm res}\equiv{\rm SNR}\sqrt{1-{\rm FF}},
\end{eqnarray} 
with the FF between the GR and MG waveforms defined as Eq.~(\ref{FF}).
$x$ is a normal random variable with zero mean and unit variance, which encodes the dependence on the noise realization.
The normalization constant ${\cal N}$ is the same in ${\cal O}_{\rm GR}$ and ${\cal O}_{\rm MG}$. 
This constant is a function of priors $P({\rm MG})$ and $P({\rm GR})$, the estimation errors, 
and the prior density widths for the MG parameters.
They cancel out when one computes $P_E$ as a function of $P_F$.

Combining Eqs.~(\ref{oddsGR}) and~(\ref{oddsMG}) with the definitions of $P_F$ and $P_E$,
one obtains
\begin{eqnarray}
 P_E=1-\frac{1}{2}( {\rm erf}(-{\rm SNR}_{\rm res}+{\rm erfc}^{-1}(P_F)) 
 -{\rm erf}(-{\rm SNR}_{\rm res}-{\rm erfc}^{-1}(P_F)) ). \label{eq:Effi}
\end{eqnarray}
The solution of (\ref{eq:Effi}), with given $P_E$ and $P_F$, is denoted as 
${\rm SNR}_{\rm res}={\rm SNR}_{\rm res}^{*}$.
A residual with ${\rm SNR}_{\rm res} \ge {\rm SNR}_{\rm res}^{*}$ is detectable, and
the SNR required for confident detection of MG corrections to GR is then simply given as a function of FF as 
${\rm SNR}_{\rm req}={\rm SNR}_{\rm res}^{*}/\sqrt{1-{\rm FF}}$.
${\rm SNR}_{\rm req}$ is the SNR required for discrimination of modified gravity models as a function of FF.
MG detectability is improved for larger SNR or for larger MG corrections, corresponding to smaller FF.
In this work, we adopt $P_E=1/2$ and $P_F=10^{-4}$, which is an appropriate value for the tens of detections expected from advanced ground-based GW detectors.
For $P_E=1/2$ and $P_F=10^{-4}$, ${\rm SNR}_{\rm req}=2.75/\sqrt{1-{\rm FF}}$.

This analysis is valid for large SNR signals, small deviations from GR, and Gaussian detector noise.

\section{Results: detectable regions in ppE parameters}
\label{sec:result}
In this section, we show what magnitude of deviations from GR waveforms are detectable 
by a single detection scenario using an advanced ground-based GW detector.
We evaluate the detectable region of the ppE magnitude parameter $\alpha$ ($\beta$) as a function of PN order 
rather than powers of the frequency $a$ ($b$). 
Here, we assume the detection threshold ${\rm SNR}=8$.
We consider three cases of CCBs as GW sources:
BNS with $(1.4~\solM,~1.4~\solM)$ [$f_{\rm ISCO}=1570$~Hz],
NSBH with $(1.4~\solM,~15~\solM)$ [$f_{\rm ISCO}=268$~Hz], 
low-mass BBH with $(8~\solM,~15~\solM)$ [$f_{\rm ISCO}=191$~Hz],
and high-mass BBH with $(30~\solM,~30~\solM)$ [$f_{\rm ISCO}=$73.28~Hz].\footnote{According to \cite{Kinugawa:2014zha,Kinugawa:2015nla}, typical masses for Pop III BBHs are $(m_1,~m_2)=(30~\solM,~30~\solM)$.}

We obtain ${\rm SNR}_{\rm req}$ from Eq.~(\ref{eq:Effi}) by setting $P_E=1/2$ and $P_F=10^{-4}$.
The detectable region of the ppE corrections is the region where ${\rm SNR}>8$ and ${\rm SNR} > {\rm SNR}_{\rm req}$ is satisfied.

\begin{figure}[h]
 \begin{minipage}[b]{0.5\linewidth}
  \centering
   \includegraphics[width=0.9\textwidth,angle=0]{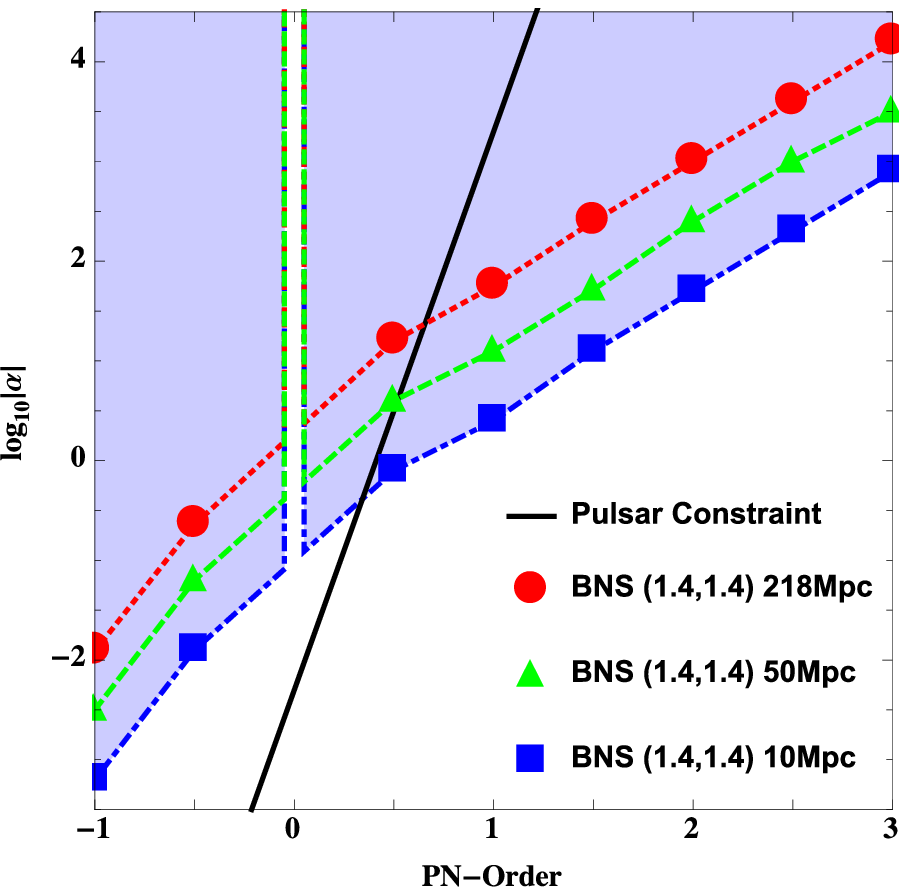}
 \end{minipage}
 \begin{minipage}[b]{0.5\linewidth}
  \centering
   \includegraphics[width=0.9\textwidth,angle=0]{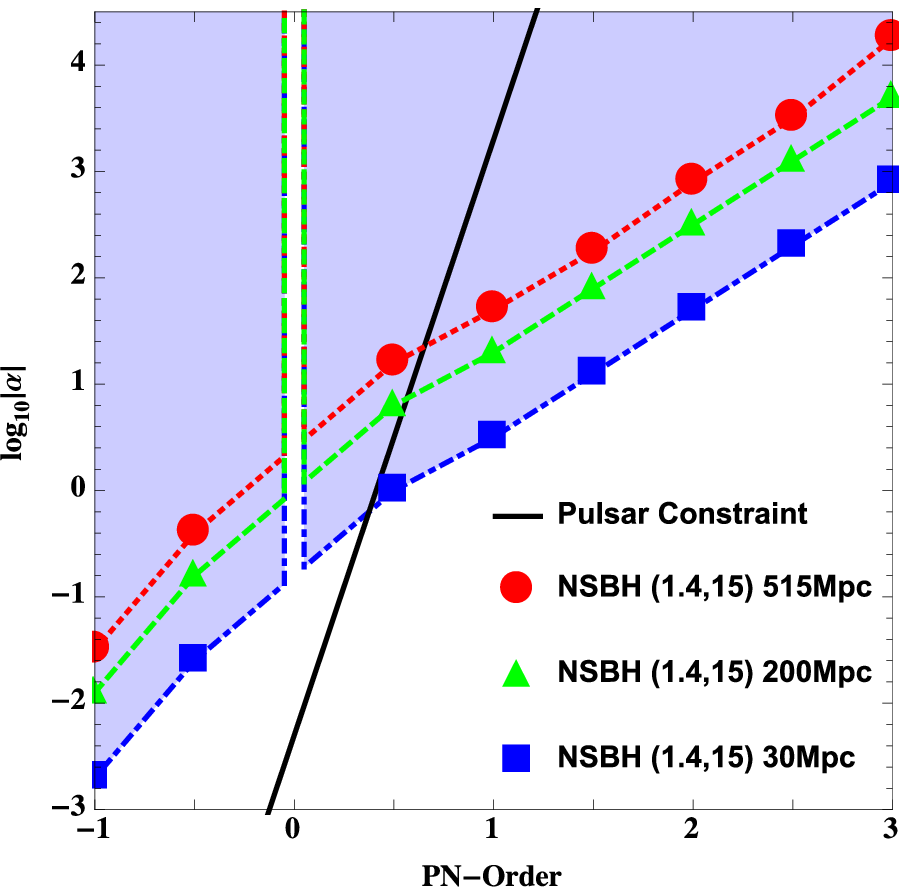}
 \end{minipage}\\
 \begin{minipage}[b]{0.5\linewidth}
  \centering
   \includegraphics[width=0.9\textwidth,angle=0]{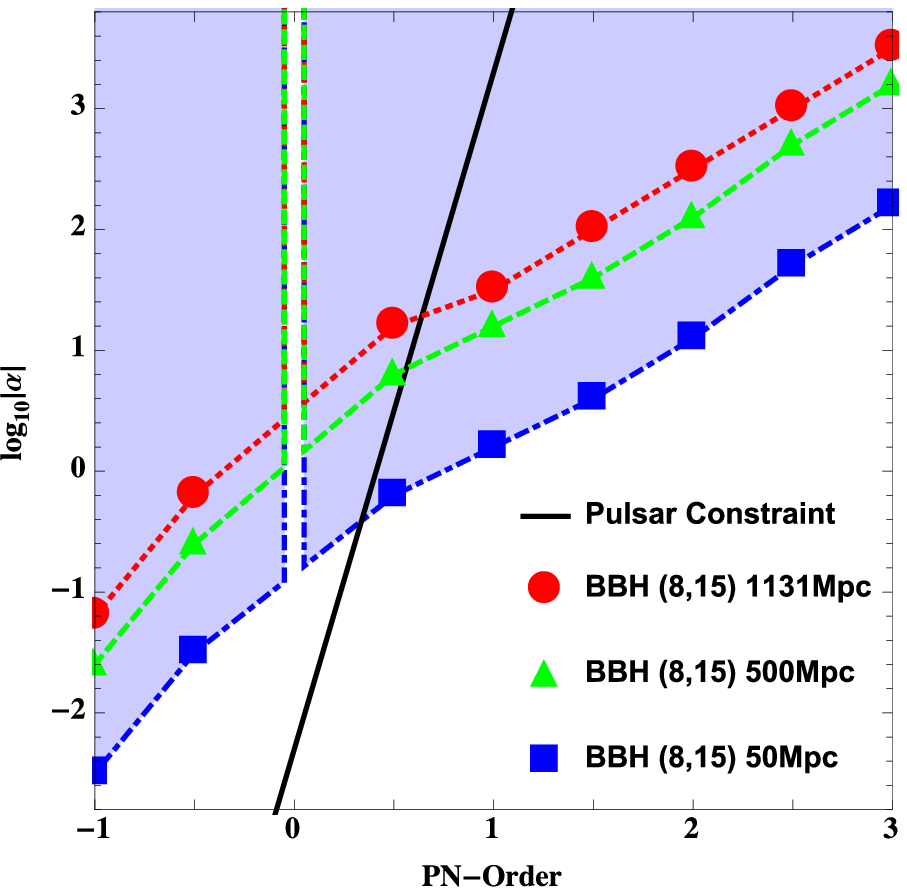}
 \end{minipage}
 \begin{minipage}[b]{0.5\linewidth}
  \centering
   \includegraphics[width=0.9\textwidth,angle=0]{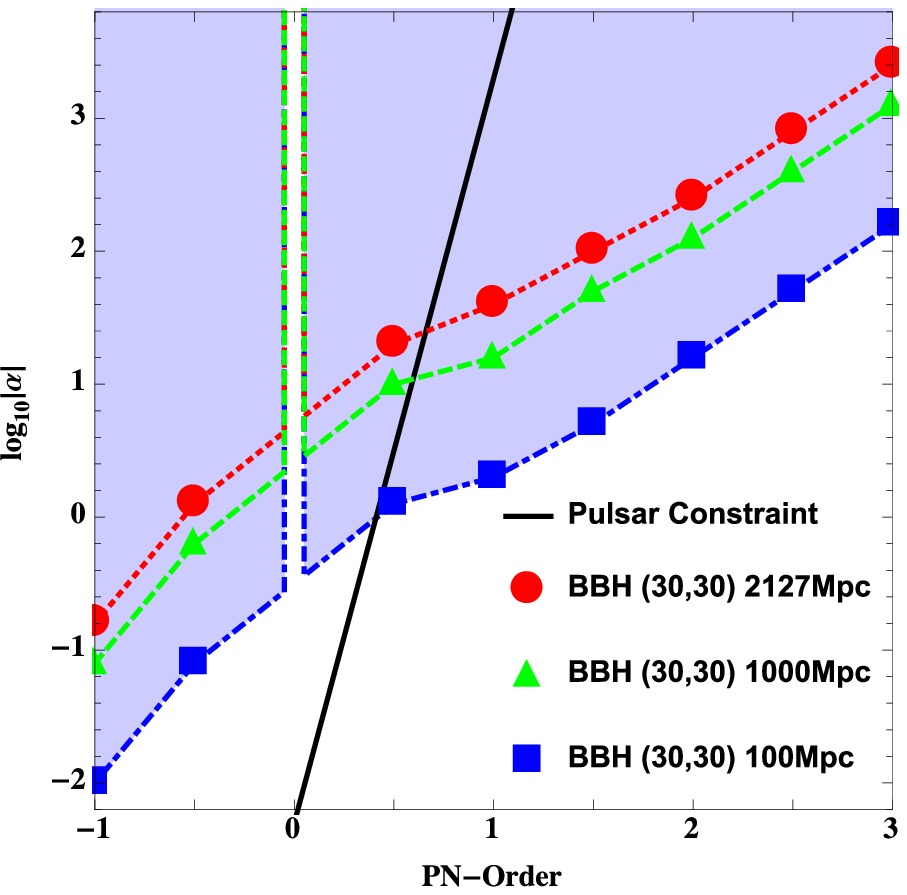}
 \end{minipage}
 \caption{
{\small
The detectable regions of the ppE amplitude parameter $\alpha$ as a function of PN order 
where ${\rm SNR}>{\rm SNR}_{\rm req}$ is satisfied.
The upper-left panel is for BNS with $(1.4~\solM,~1.4~\solM)$.
Each region above each mark/curve is the detectable region for different distance to the source.
Shaded regions are the detectable region for an extremely loud event at $D_L=10~{\rm Mpc}$.
Here, the detection efficiency is set to $P_E=1/2$, and the false alarm probability is set to $P_F=10^{-4}$.
Each distance corresponds to ${\rm SNR}=8$ ($D_L=218~{\rm Mpc}$), 
${\rm SNR}=34.9$ ($D_L=50~{\rm Mpc}$), and ${\rm SNR}=174.6$ ($D_L=10~{\rm Mpc}$), 
in the case of GR respectively.
The solid line is the bound from binary pulsar observations 
of orbital period decay due to GW emission for ${\rm PSR~J0737-3039}$.
The regions below this bounds have been not yet excluded.
The upper-right panel is a plot similar to the upper-left but for NSBH with $(1.4~\solM,~15~\solM)$.
The marks/curves correspond to the distance to the source $D_L=515~{\rm Mpc}$ [circle/dotted (red)],
$200~{\rm Mpc}$ [triangle/dashed (green)], and $30~{\rm Mpc}$ [square/dot-dashed (blue)].
Each distance corresponds to ${\rm SNR}=8$ ($D_L=515~{\rm Mpc}$), 
${\rm SNR}=20.6$ ($D_L=200~{\rm Mpc}$), and ${\rm SNR}=137.2$ ($D_L=30~{\rm Mpc}$), 
in the case of GR respectively.
The lower-left panel is a plot similar to the upper-left but for low-mass BBH with $(8~\solM,~15~\solM)$.
The marks/curves correspond to the distance to the source $D_L=1131~{\rm Mpc}$ [circle/dotted (red)],
$500~{\rm Mpc}$ [triangle/dashed (green)], and $50~{\rm Mpc}$ [square/dot-dashed (blue)].
Each distance corresponds to ${\rm SNR}=8$ ($D_L=1131~{\rm Mpc}$), 
${\rm SNR}=18.1$ ($D_L=500~{\rm Mpc}$), and ${\rm SNR}=181.0$ ($D_L=50~{\rm Mpc}$), 
in the case of GR respectively.
The lower-right panel is a plot similar to the upper-left but for high-mass BBH with $(30~\solM,~30~\solM)$.
The marks/curves correspond to the distance to the source $D_L=2127~{\rm Mpc}$ [circle/dotted (red)],
$1000~{\rm Mpc}$ [triangle/dashed (green)], and $100~{\rm Mpc}$ [square/dot-dashed (blue)].
Each distance corresponds to ${\rm SNR}=8$ ($D_L=2127~{\rm Mpc}$), 
${\rm SNR}=17.0$ ($D_L=1000~{\rm Mpc}$), and ${\rm SNR}=170.1$ ($D_L=100~{\rm Mpc}$), 
in the case of GR respectively.
}
 }\label{fig:amp}
\end{figure}

\subsection{Amplitude corrections}
\label{sec:amplitudecorrection}
First, we consider the ppE corrections only in the amplitude as expressed in Eq.~(\ref{hppEamp}). We treat the ppE parameters 
$\alpha$ and $a$ as independent parameters.
We assess what magnitude of the ppE amplitude corrections $\alpha$ is detectable
as a function of PN order by using Vallisneri's method.

Figure~\ref{fig:amp} shows the detectable regions of the ppE amplitude corrections $|\alpha|$ to the waveforms, 
as a function of PN order.
The upper-left panel of Fig.~\ref{fig:amp} is for BNS with $(1.4~\solM,~1.4~\solM)$.
The marks/curves correspond to the source at horizon distance $D_L=218~{\rm Mpc}$ [circle/dotted (red)],
the source at an intermediate distance, $D_L=50~{\rm Mpc}$ [triangle/dashed (green)], 
and an extremely loud event with $D_L=10~{\rm Mpc}$ [square/dot-dashed (blue)]. 
Each distance corresponds to ${\rm SNR}=8$ ($D_L=218~{\rm Mpc}$), 
${\rm SNR}=34.9$ ($D_L=50~{\rm Mpc}$), and ${\rm SNR}=174.6$ ($D_L=10~{\rm Mpc}$), 
in the case of GR respectively. 
The region above each mark/curve is the region in which the ppE corrections are detectable.
The shaded regions are the detectable regions for an extremely loud event at $D_L=10~{\rm Mpc}$.
The solid line (black) is the current bound from binary pulsar observations of orbital period decay due to GW emission
for the binary pulsar ${\rm PSR~J0737-3039}$~\cite{Yunes:2010qb}.\footnote{One needs to be careful in comparing bounds on ppE parameters from binary pulsar and GW observations. This is because typically ppE $\alpha$ and $\beta$ depend not only on theoretical coupling constants (like the Brans‐Dicke parameter) but also on system parameters like masses and spins of a binary. Thus, binary pulsar bounds on $\alpha$ and $\beta$ do not directly apply to those from GW observations.}
The region above this solid line is already excluded. 
Thus, the region below the solid line is the region where there is a possibility of detecting 
the ppE correction. 
We find that there is a possibility of detecting the amplitude correction 
at 1PN and higher order, which is not excluded by the binary pulsar observations. 
This is true even for the horizon distance events with ${\rm SNR}=8$. 

The detectability of the ppE corrections becomes weaker 
as PN order (or powers of frequency $a$) becomes higher.
At 0PN order, a GW observation cannot distinguish deviations from GR waveforms because 
the correction term, $1+\alpha$, is degenerate with the distance $D_L$. 

The detectable region becomes larger as the distance $D_L$ becomes smaller.
Notice that ${\rm SNR}$ depends on $(a, ~\alpha)$ and 
$D_{\rm L}$ for fixed component masses. 
On the other hand, 
${\rm SNR}_{\rm req}$ does not depend on $D_{\rm L}$ but depends on $(a, ~\alpha)$.
Thus, the dependence on the distance comes from the difference in ${\rm SNR}$.

We also examine the dependence of our results on masses: 
NSBH with $(1.4~\solM,~15~\solM)$ [upper-right panel of Fig.~\ref{fig:amp}],
low-mass BBH with $(8~\solM,~15~\solM)$ [lower-left panel of Fig.~\ref{fig:amp}],
and high-mass BBH with $(30~\solM,~30~\solM)$ [lower-right panel of Fig.~\ref{fig:amp}]. 
The features of the detectable region for these cases are similar to the BNS case;
that is, the ppE correction is detectable at 1PN and higher PN orders. 
The circle/dotted curves of each figure (red online) represent the case for ${\rm SNR=8}$ in GR. 
The detectable regions for fixed SNR become slightly  smaller as the total mass increases, 
but this dependence is not very large. 
The constraint on 1PN order for ${\rm SNR=8}$ case is 
$\log_{10}|\alpha| \simgt 1.75 ~(1.4~\solM,~1.4~\solM)$,
$\log_{10}|\alpha| \simgt 1.7 ~(1.4~\solM,~15~\solM)$, 
$\log_{10}|\alpha| \simgt 1.5 ~(8~\solM,~15~\solM)$, and 
$\log_{10}|\alpha| \simgt 1.6 ~(30~\solM,~30~\solM)$,
respectively.

\subsection{Phase corrections}
\label{sec:phasecorrection}
Next, we consider the ppE corrections in the phase as expressed in Eq.~(\ref{hppEphase}). We treat the ppE parameters 
$\beta$ and $b$ as independent parameters.
We assess what magnitude of the ppE phase corrections $\beta$ is detectable 
as a function of PN order.

\begin{figure}[h]
 \begin{minipage}[b]{0.5\linewidth}
  \centering
   \includegraphics[width=0.9\textwidth,angle=0]{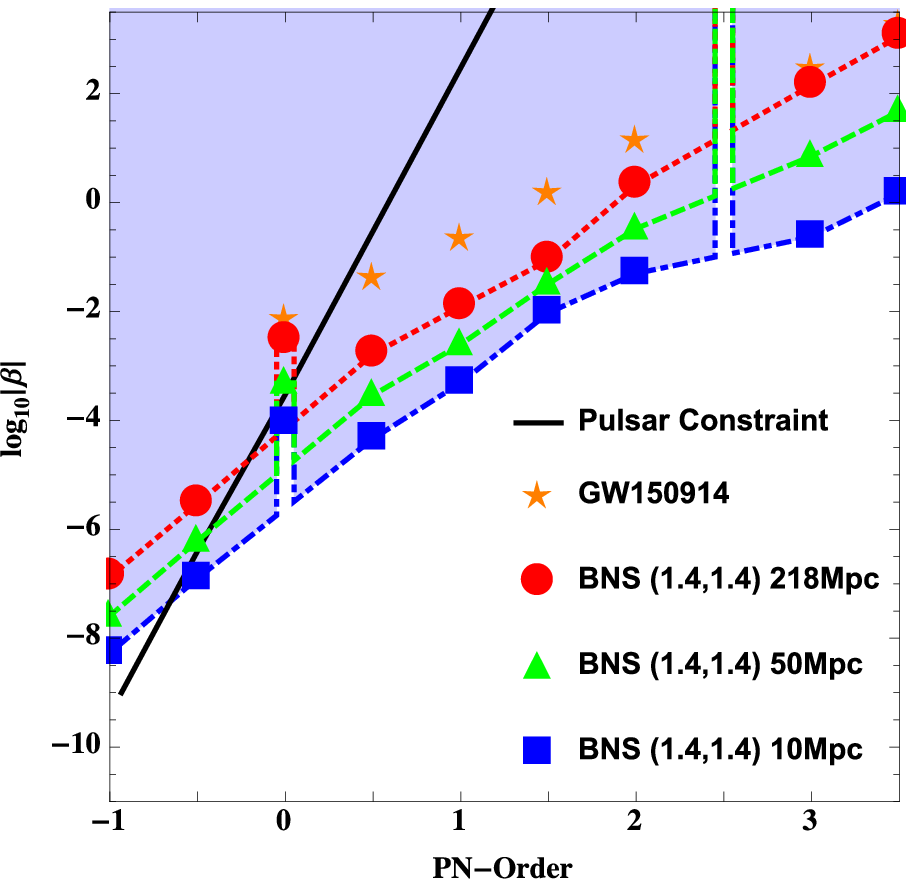}
 \end{minipage}
 \begin{minipage}[b]{0.5\linewidth}
  \centering
   \includegraphics[width=0.9\textwidth,angle=0]{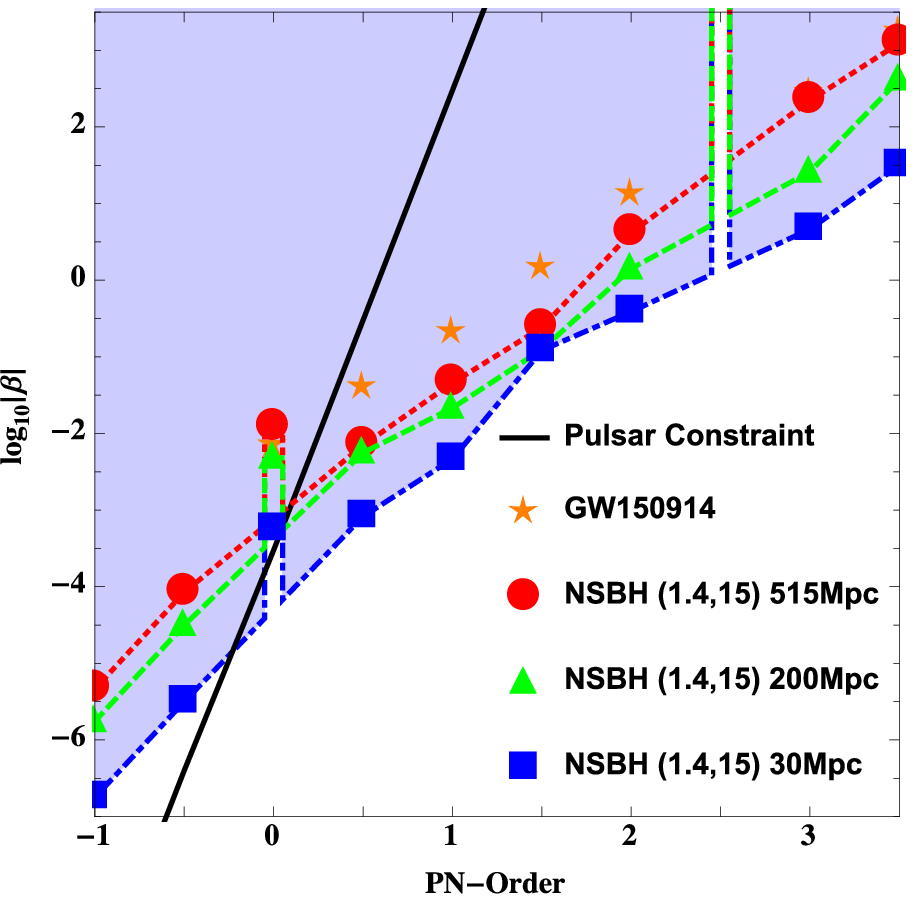}
 \end{minipage}\\
 \begin{minipage}[b]{0.5\linewidth}
  \centering
   \includegraphics[width=0.9\textwidth,angle=0]{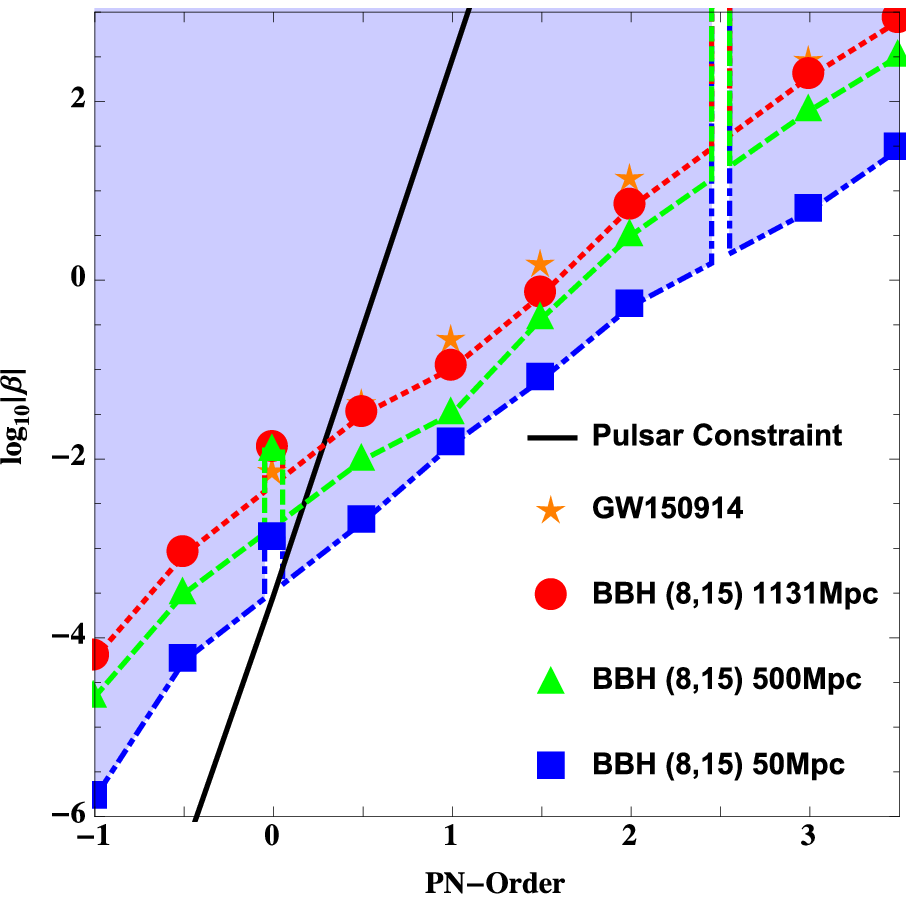}
 \end{minipage}
 \begin{minipage}[b]{0.5\linewidth}
  \centering
   \includegraphics[width=0.9\textwidth,angle=0]{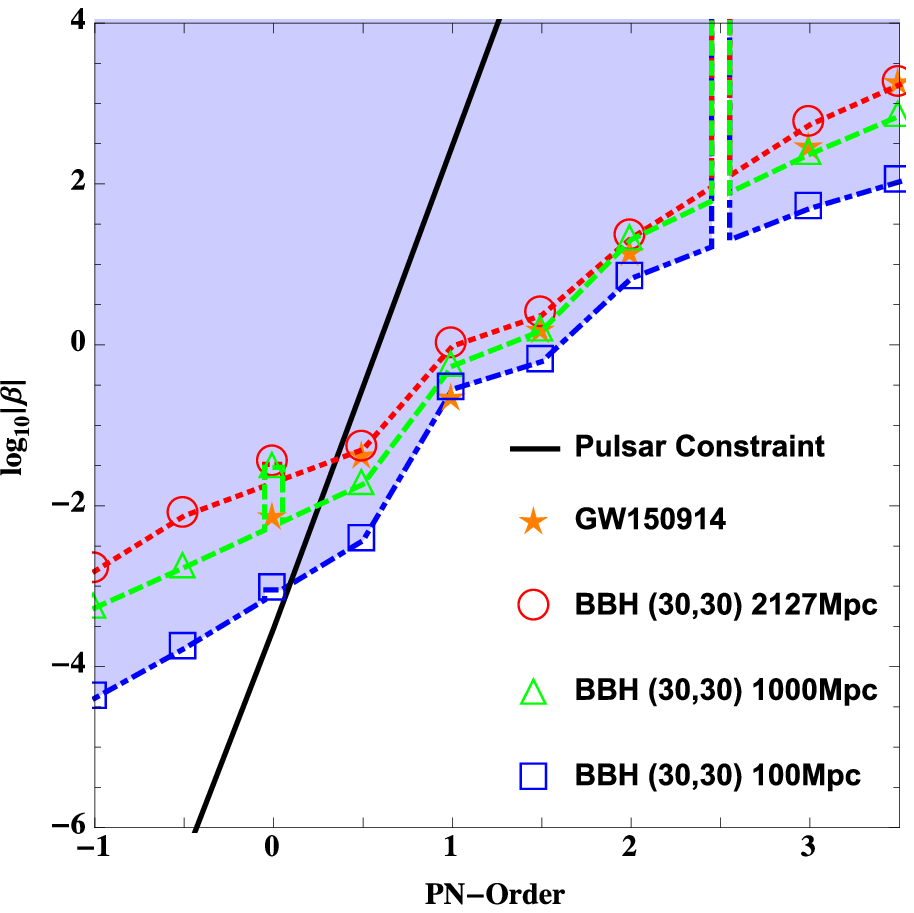}
 \end{minipage}
 \caption{
 {\small
 The detectable regions of the ppE phase parameter $\beta$ as a function of PN order  
where ${\rm SNR}>{\rm SNR}_{\rm req}$ is satisfied.
The upper-left panel is for BNS with $(1.4~\solM,~1.4~\solM)$.
The definitions of the marks/curves are the same as those of Fig.~\ref{fig:amp}.
Each distance corresponds to ${\rm SNR}=8$ ($D_L=218~{\rm Mpc}$), 
${\rm SNR}=34.9$ ($D_L=50~{\rm Mpc}$), and ${\rm SNR}=174.6$ ($D_L=10~{\rm Mpc}$), 
respectively.
The orange stars represent the bounds from GW150914~\cite{TheLIGOScientific:2016src,Yunes:2016jcc}.
The upper-right panel is a plot similar to upper-left panel but for NSBH with $(1.4~\solM,~15~\solM)$.
The definitions of the marks/curves are the same as those of the upper-right panel of Fig.~\ref{fig:amp}.
Each distance corresponds to ${\rm SNR}=8$ ($D_L=515~{\rm Mpc}$), 
${\rm SNR}=20.6$ ($D_L=200~{\rm Mpc}$), and ${\rm SNR}=137.2$ ($D_L=30~{\rm Mpc}$), respectively.
The lower-left panel is a plot similar to upper-left panel but for low-mass BBH with $(8~\solM,~15~\solM)$.
The definitions of the marks/curves are the same as those of the lower-left panel of Fig.~\ref{fig:amp}.
Each distance corresponds to ${\rm SNR}=8$ ($D_L=1131~{\rm Mpc}$), 
${\rm SNR}=18.1$ ($D_L=500~{\rm Mpc}$), and ${\rm SNR}=181.0$ ($D_L=50~{\rm Mpc}$), respectively.
The lower-right panel is a plot similar to the upper-left panel but for high-mass BBH with $(30~\solM,~30~\solM)$.
The definitions of the marks/curves are the same as those of the lower-right panel of Fig.~\ref{fig:amp}.
Each distance corresponds to ${\rm SNR}=8$ ($D_L=2127~{\rm Mpc}$), 
${\rm SNR}=17.0$ ($D_L=1000~{\rm Mpc}$), and ${\rm SNR}=170.1$ ($D_L=100~{\rm Mpc}$), 
respectively.
}
 }\label{fig:phase_PP}
\end{figure}

Figure~\ref{fig:phase_PP}
shows the detectable regions of the ppE phase corrections $\beta$ to GR waveforms 
as a function of PN order 
for non-spinning and point-particle binaries with a circular orbit.
The upper-left panel of Fig.~\ref{fig:phase_PP} is for BNS with $(1.4~\solM,~1.4~\solM)$.
The meaning of this figure is the same as that of Fig.~\ref{fig:amp}.  
We find that there is the possibility of detecting the phase correction at 0.5PN and higher orders,
which is not excluded by binary pulsar observations. 
The orange stars represent the bounds from GW150914~\cite{TheLIGOScientific:2016src,Yunes:2016jcc}, 
which was produced by a binary black hole of masses around $(35~\solM,~30~\solM)$ at the source frame and $D_L \simeq 400~{\rm Mpc}$~\cite{TheLIGOScientific:2016wfe,Abbott:2016izl}.
The detectability of the ppE corrections in the phase becomes weaker as the PN order increases. 
At 0PN order, $\beta$ is partially degenerate with the other parameters, and the constraint on $\beta$ becomes weaker.
At 2.5~PN order, corresponding to $b=0$, since the correction $\beta$ is completely degenerate with the phase $\Phi_0$,
it is not possible to distinguish deviations from GR waveforms from $\Phi_0$. 
Thus, there is no constraint at this order. 

We also examine the dependence of our results on component masses for 
NSBH with $(1.4~\solM,~15~\solM)$ [upper-right panel of Fig.~\ref{fig:phase_PP}],
low-mass BBH with $(8~\solM,~15~\solM)$ [lower-left panel of Fig.~\ref{fig:phase_PP}],
and high-mass BBH with $(30~\solM,~30~\solM)$ [lower-right panel of Fig.~\ref{fig:phase_PP}].
We find that, for all of these cases, 
there is a detectable region at 0.5PN  and higher order, which is not excluded by binary pulsar observations. 
There is a tendency that the detectable region becomes smaller for larger mass cases. 
This is because, for fixed ${\rm SNR}$, 
the number of cycles ${\cal N}_{\rm cyc}$ spent in the interferometers' band
decreases as the chirp mass increases. 
Thus it becomes more difficult to detect the small changes of the coefficients of 
PN terms for larger total mass cases. 
However, the mass dependence of the detectable region is not very large. 
The constraint on 1PN order for ${\rm SNR=8}$ case is 
$\log_{10}|\beta| \simgt -1.84 ~(1.4~\solM,~1.4~\solM)$,
$\log_{10}|\beta| \simgt -1.34 ~(1.4~\solM,~15~\solM)$, 
$\log_{10}|\beta| \simgt -0.98 ~(8~\solM,~15~\solM)$, and 
$\log_{10}|\beta| \simgt -0.017 ~(30~\solM,~30~\solM)$,
respectively.

We also investigate whether the spin effects on our results are consistent with those in some previous works~\cite{Agathos:2013upa,Sampson:2013lpa,TheLIGOScientific:2016src,Yunes:2016jcc}.
What is new in this paper compared to the relevant previous works is the study of the effects due to orbital eccentricity or tidal deformation on detectable regions of generic deviations from GR waveforms.

\begin{figure}[h]
 \begin{minipage}[b]{0.5\linewidth}
  \centering
   \includegraphics[width=0.9\textwidth,angle=0]{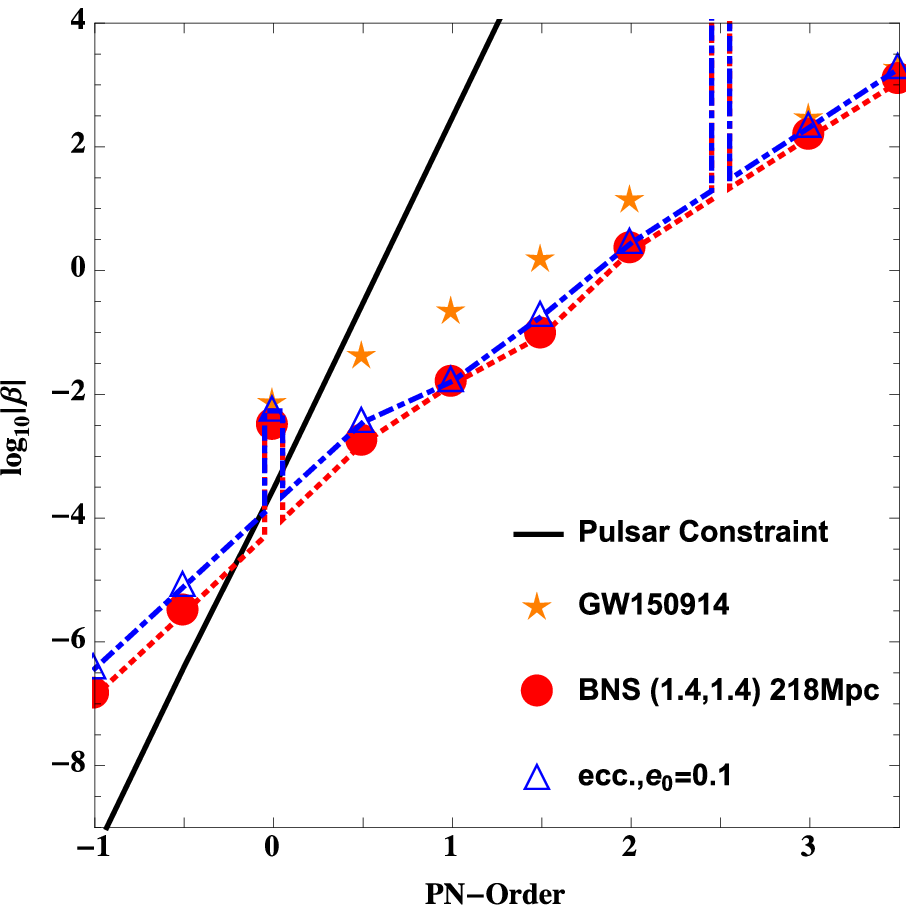}
 \end{minipage}
 \begin{minipage}[b]{0.5\linewidth}
  \centering
   \includegraphics[width=0.9\textwidth,angle=0]{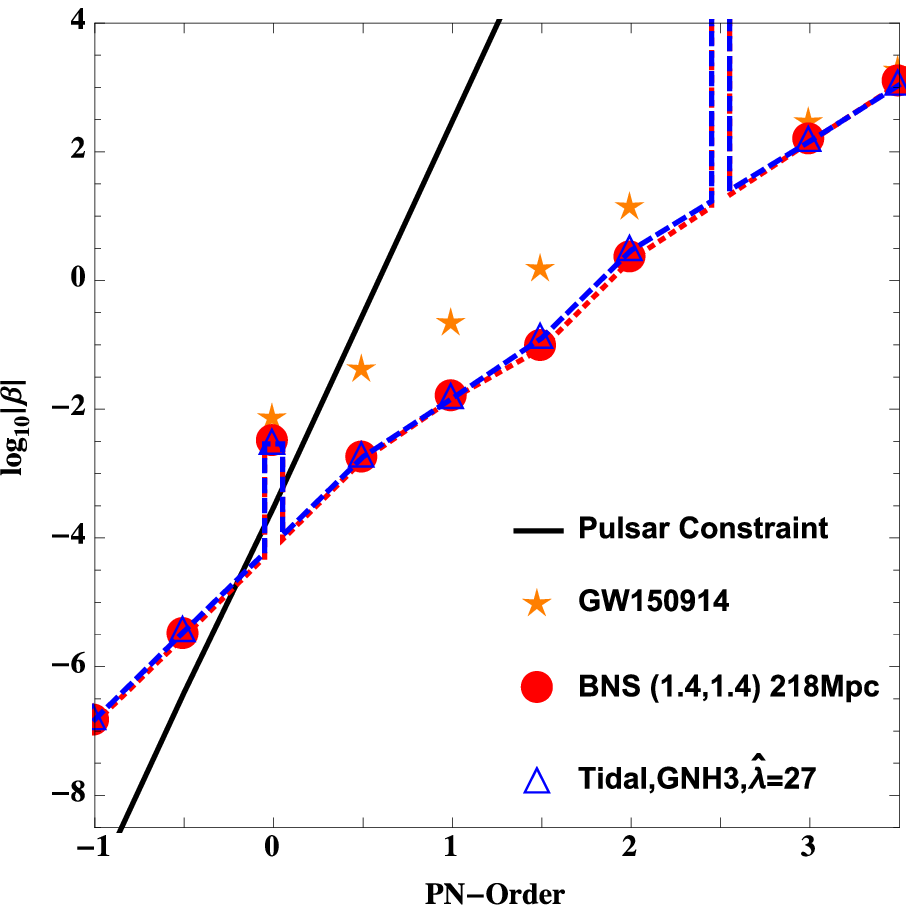}
 \end{minipage}
 \caption{
The left panel shows 
the effects of orbital eccentricity on the detectable regions of the ppE phase parameter $\beta$ as a function of PN order for BNS 
with $(1.4~\solM,~1.4~\solM)$, 
where ${\rm SNR}>{\rm SNR}_{\rm req}$ is satisfied.
The region above the  [triangle/dod-dashed (blue)]
mark/curve is the detectable region for orbital eccentricity $e_0=0.1$ at a reference frequency $f_0=10~{\rm Hz}$.
For comparison, the detectable region for the point-particle case [circle/dotted (red)] and bounds from GW150914 [stars (orange)] are shown.
The right panel shows the effects of tidal deformability on the detectable regions of the ppE phase parameter $\beta$ as a function of PN order for BNS 
with $(1.4~\solM,~1.4~\solM)$, 
where ${\rm SNR}>{\rm SNR}_{\rm req}$ is satisfied.
The region above the [triangle/dashed (blue)]
mark/curve is the detectable region for the GNH3 equation of state
with tidal deformability $\hat{\lambda}=27$.
 }\label{fig:phaseSys_BNS}
\end{figure}

The left panel of Fig.~\ref{fig:phaseSys_BNS} shows the effect of orbital eccentricity on the detectable region of the ppE phase parameter $\beta$ as a function of PN order.
We find that the systematic bias on the detectable region of $\beta$ due to orbital eccentricity is not significant even for $e_0=0.1$. 
The eccentricity has a larger effect on relatively lower PN order modifications, as the eccentricity first enters at negative PN order.

The right panel of Fig.~\ref{fig:phaseSys_BNS} shows the effect of tidal deformation on the detectable region of the ppE phase parameter $\beta$ as a function of PN order.
We consider the GNH3 equation of state~\cite{Glendenning:1984jr} for NSs listed in Table I of ~\cite{Damour:2012yf},
 which yields a configuration with small compactness and would have given a relatively large value of $\hat{\lambda}$.
The GNH3 equation of state was derived by using the mean field approximation for the Lagrangian of interacting nucleons, hyperons, and mesons. 
In this analysis, each integral ~(\ref{inner_product}) is taken from $f_{\rm max}=f_{\rm contact}$, where $f_{\rm contact}$ is the gravitational wave frequency at contact and is $1284~{\rm Hz}$ for GNH3.
We find that the systematic bias on the detectable region of $\beta$ due to tidal deformation is not significant even for 
the GNH3 equation of state with $\hat{\lambda}=27$.
The tidal effects have a larger effect on relatively higher PN modifications, as they first enter at 5PN order.
However, since the effects are small even for relatively higher PN orders,
it is difficult to see such features in the right panel of Fig.~\ref{fig:phaseSys_BNS}.

\section{Summary and discussion}
\label{sec:summary}
We examined the detectable region of generic deviations from GR 
for four representative coalescing compact binaries:
BNS, NSBH, low-mass BBH, and high-mass BBH.
Our results show that advanced ground-based gravitational wave detectors have the potential to detect or constrain 
generic deviations from GR waveforms
in the interesting parameter regions which are not excluded by binary pulsar observations.
The constraints on the amplitude coefficient, $\alpha$, 
at 1PN order for ${\rm SNR=8}$ case are 
$\log_{10}|\alpha| \simgt 1.75 ~(1.4~\solM,~1.4~\solM)$ and
$\log_{10}|\alpha| \simgt 1.5 ~(8~\solM,~15~\solM)$, 
respectively. 
The constraints on the phase coefficient, $\beta$, 
at 1PN order for ${\rm SNR=8}$ case are
$\log_{10}|\beta| \simgt -1.84 ~(1.4~\solM,~1.4~\solM)$ and
$\log_{10}|\beta| \simgt -0.98 ~(8~\solM,~15~\solM)$, 
respectively.
The detectable region evaluated here reflects the effects of both statistical and 
systematic biases on model selection based on Bayesian statistics.

We also show that the systematic bias on the detectable region of the ppE phase parameter $\beta$ due to orbital eccentricity or tidal deformation
is not significant even for $e_0=0.1$ or $\hat{\lambda}=27$, respectively.
We assume that phase corrections for spin, orbital eccentricity, and tidal deformation to the ppE waveform 
are the same expression (\ref{phase_sys_ppE}) as those of GR (\ref{phase_sys_GR}).
However, for consistency, we have to use waveforms derived from the energy balance equation of modified gravity corrected  by systematic effects.
We plan to investigate the systematic effects for modified gravity by using such waveforms in the future.

In this work, we assumed that the distance to the source is known.
In real data analysis, it is possible to determine the distance as well as the direction to the source 
and the inclination angle by using a network of GW detectors. 
Electromagnetic follow-up observations could also be helpful. 
We will investigate the detectability of MG effects with networks of detectors
in the future. 

We have not included the spin angular momenta of the compact stars. 
If the spin precession effect exists, there will be an amplitude modulation. 
Such modulation may affect the modification caused by MG effects.
In this work, we consider the case of a single detection. 
However, if several GW events are detected, 
more stringent constraints can be obtained. 
We leave these issues as topics to be investigated in the future.

\section*{Acknowledgments}
This work was supported by MEXT Grant-in-Aid for Scientific Research on Innovative Areas, 
``New Developments in Astrophysics Through Multi-Messenger Observations of Gravitational Wave Sources''
Grant No. 24103005.
This work was also supported by JSPS KAKENHI Grant Numbers 
23540309, 24244028, and 15K05081, and 
by the JSPS Core-to-Core Program, A. Advanced Research Networks.



\end{document}